\documentclass[pra,aps,twocolumn,10pt,showpacs,groupedaddress,superscriptaddress,floatfix,notitlepage]{revtex4-1}
\usepackage{amsmath,amsfonts,amssymb,graphics,graphicx,epsfig,color,times}
\usepackage[utf8x]{inputenc}
\usepackage{color}
\usepackage{bbm, dsfont} 
\usepackage{subfigure}
\usepackage{hyperref}
\usepackage{mathrsfs}
\usepackage{verbatim}
\begin{document}
\title{Subradiance dynamics in a singly-excited chiral-coupled atomic chain}
\author{H. H. Jen}
\email{sappyjen@gmail.com}
\affiliation{Institute of Physics, Academia Sinica, Taipei 11529, Taiwan}
\author{M.-S. Chang}
\affiliation{Institute of Atomic and Molecular Sciences, Academia Sinica, Taipei 10617, Taiwan}
\affiliation{Center for Quantum Technology and Department of Physics, National Tsing Hua University, Hsinchu 30013, Taiwan}
\author{G.-D. Lin}
\affiliation{CQSE, Department of Physics, National Taiwan University, Taipei 10617, Taiwan}
\author{Y.-C. Chen}
\affiliation{Institute of Atomic and Molecular Sciences, Academia Sinica, Taipei 10617, Taiwan}

\date{\today}
\renewcommand{\r}{\mathbf{r}}
\newcommand{\f}{\mathbf{f}}
\renewcommand{\k}{\mathbf{k}}
\def\p{\mathbf{p}}
\def\q{\mathbf{q}}
\def\bea{\begin{eqnarray}}
\def\eea{\end{eqnarray}}
\def\ba{\begin{array}}
\def\ea{\end{array}}
\def\bdm{\begin{displaymath}}
\def\edm{\end{displaymath}}
\def\red{\color{red}}
\pacs{}
\begin{abstract}
We theoretically investigate the subradiance dynamics in a nonreciprocal chiral-coupled atomic chain, in which infinite-range dipole-dipole interaction emerges in the dissipation. We find that super- and subradiance are both present in the dissipation process following single photon excitation, and the decay dynamics shows burst emissions from uniform initial excitations, which reflects the influence of atomic ordering on the propagation of light-induced atom-atom correlations. By tuning the nonreciprocal couplings in the chiral-coupled atomic system, we show that the subradiance dynamics can be greatly modified. We further study the effect of atomic local disorder, and find occurrence of plateaus on the decay curve dependent on the defect locations, as well as persistent localized excitations induced by disorders. We also discuss the effect of imperfections of systems on the subradiance dynamics. Our results show rich opportunities in the chiral-coupled system toward photon storage and routing. 
\end{abstract}
\maketitle
\section{Introduction}

Chiral quantum optics \cite{Lodahl2017}, a new paradigm of controlling light-matter interactions in one-dimensional (1D) nanophotonics devices \cite{Chang2018}, opens up many opportunities in quantum information processing and quantum simulation. Chiral coupling allows a nonreciprocal atom-light interface, which breaks the time-reversal symmetry in conventional light-matter interacting systems in free space. This nonreciprocal coupling emerges due to the strong radial confinement of 1D reservoirs which couples the atoms and spontaneously radiates only in the guided dimension \cite{Tudela2013}. The chiral-coupled interface can be realized in the setting of evanescent waves \cite{Bliokh2014, Bliokh2015} at the glass-air surface under total internal reflection. Due to considerable reduction of the evanescent waves in the normal direction of the surface, the longitudinal component of light becomes finite and enables the transverse spin angular momentum that can be locked to the propagation direction. This spin-momentum locking or spin-orbital coupling is the essential element in the chiral-coupled systems.   

There are important applications in such 1D atom-light interacting systems. In an atom-fiber coupled system, directional spontaneous emissions can be controlled by the internal states of the atoms \cite{Mitsch2014}. In an atom-resonator system, photon routing can be achieved by single-atom switch controlled by the single photon, in which a reflection of the photon toggles the switch from reflection to high transmission \cite{Shomroni2014}. Other than these atom-light interfaces, quantum dot displaced from the crossing region of in-plane nanowire waveguides \cite{Luxmoore2013} can guide photons and provide an interface between a solid-state spin qubit and path-encoded photons. Similarly in the setting of quantum dot in the glide-plane photonic crystal waveguide \cite{Sollner2015} under external magnetic field, chiral coupling of the system realizes a Mach-Zehnder interferometer. Even a CNOT gate can potentially be implemented in such systems \cite{Sollner2015}. In the setting of two superconducting qubits in a 1D waveguide, nonreciprocal coupling can also be realized via quantum nonlinear couplings under quasi-dark state \cite{Hamann2018}.

Recently, protocols of quantum state transfer using chiral-coupled photonic quantum link were theoretically proposed \cite{Grankin2018}, and selective transport of atomic excitations can be achieved in a driven chiral-coupled atomic chain \cite{Jen2019_driven}. Chiral-coupled 1D atomic chain can even realize quantum many-body states of spin dimers \cite{Stannigel2012, Ramos2014, Pichler2015} or simulate exotic photonic topological quantum states \cite{Lodahl2017}. On the recent progress of 1D reciprocally-coupled atom-light systems, protocols were proposed to create mesoscopic entangled states by engineering the collective decay dynamics \cite{Tudela2013}, which revealed emerging universal behavior in the coherent dynamics, i.e., collective frequency shift of the resonant dipole-dipole interaction, of the systems \cite{Kumlin2018}.

\begin{figure}[t]
\centering
\includegraphics[width=8.5cm,height=6.5cm]{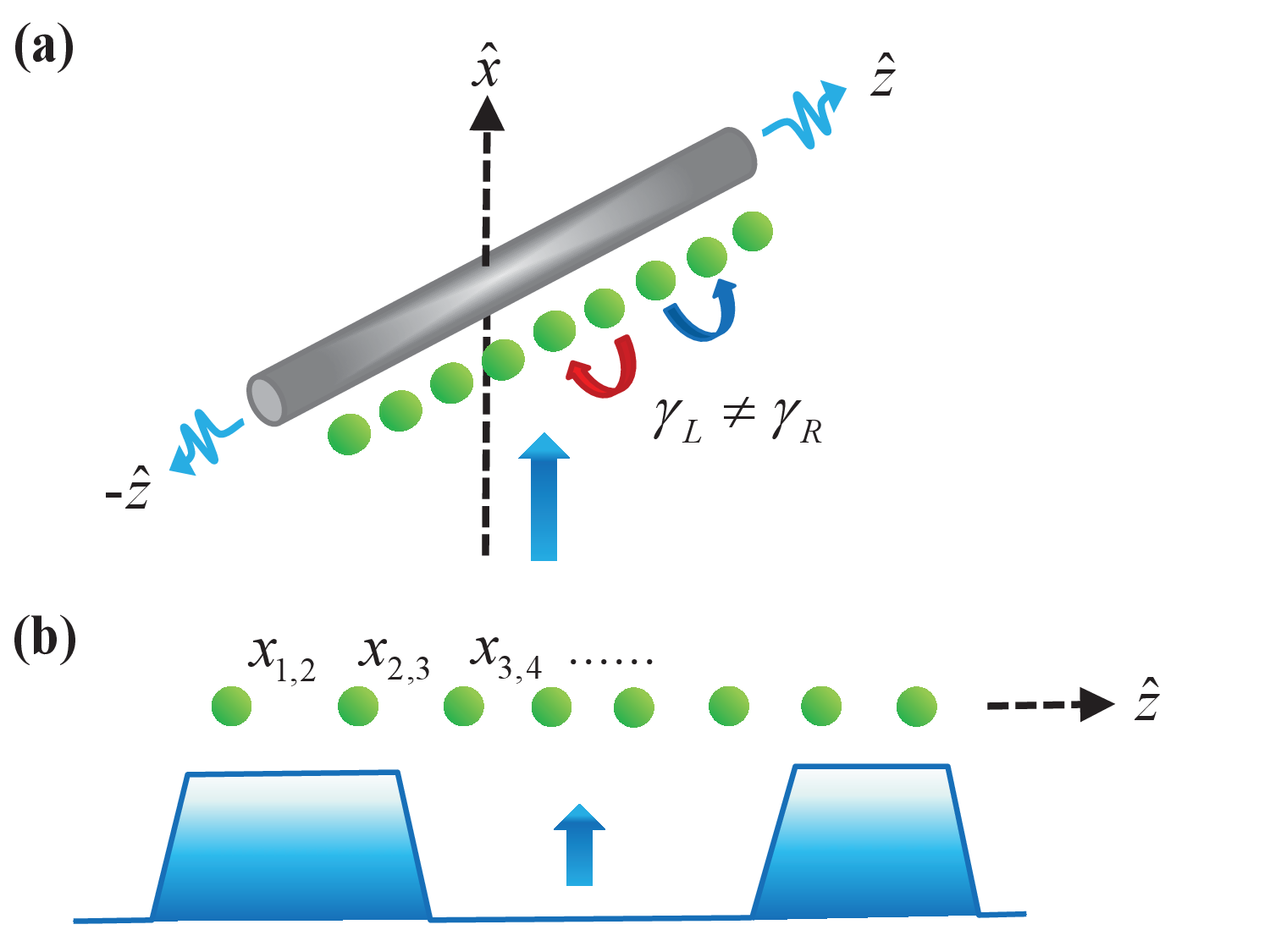}
\caption{Chiral-coupled atomic chain. (a) Schematic one-dimensional atom-fiber coupled system demonstrates one example of the chiral coupling. Single photon propagating in $\hat x$ excites the atomic chain distributed along $\hat z$ near the fiber, and the atom-fiber coupled system guides the spontaneous emissions with the nonreciprocal couplings $\gamma_L\neq\gamma_R$. (b) Illustrative structured beams which non-uniformly excite the atomic chain. Controlled beam profiles and/or inter-atomic separations $x_{1,2}$, $x_{2,3}$, $x_{3,4}$, $...$, allow tailoring the state initializations which modify the cooperative couplings.}\label{fig1}
\end{figure}

In this article, we investigate the dynamics of the spontaneous emissions in the 1D chiral-coupled atomic chain as in Fig. \ref{fig1}, since clear identification of super- \cite{Dicke1954, Lehmberg1970, Gross1982, Jen2012, Jen2015, Araujo2016, Roof2016, Jennewein2016, Bromley2016, Zhu2016, Shahmoon2017} and sub-radiance \cite{Scully2015, Guerin2016, Facchinetti2016, Jen2016_SR, Sutherland2016, Bettles2016, Jen2017_MP, Jenkins2017, Garcia2017, Jen2018_SR1, Jen2018_SR2} in such system is less studied. Additionally, since the 1D reservoir allows nonreciprocal decay channels and can be tailored via manipulation of atomic separations and/or excitation beam profiles, the cooperative radiation along the allowed dimension should possess qualitatively different features compared to that in reciprocally-coupled systems.

The rest of the paper is organized as follows. In Sec. II, we obtain the coupling matrix for a 1D chiral-coupled atomic chain with single excitation, and analyze few-atom cases. In Sec. III, we characterize the subradiance for longer atomic chains. In Sec. IV, we further study the effect of dislocations of the atoms on the radiation properties. Finally in Sec. V we discuss the effect of imperfections of systems and conclude in Sec. VI. In the Appendix, we review the general formalism for resonant dipole-dipole interaction (RDDI) of the spontaneous emissions in 1D, two-dimensional (2D), and three-dimensional (3D) reservoirs \cite{Lehmberg1970}. 

\section{Chiral coupling matrix in single excitation Hilbert space}

Conventional light-matter interacting systems does not constrain which direction spontaneous emission should radiate into; therefore, the RDDI symmetrically couples every pair of atoms in the system, and we should have the reciprocal form of RDDI, which preserves the time reversal symmetry of light scattering, and this property should also preserve in 1D and 2D spaces of reservoirs. 

In contrast to the RDDI which we generally obtain and review in the Appendix, the chiral-coupled system allows nonreciprocal decay channels which break the time reversal symmetry. The effective chiral master equation of 1D coupled atom-light interacting system in Lindblad forms \cite{Pichler2015} gives
\bea
\frac{d Q}{dt}&&=-\frac{i}{\hbar}[ Q, H_L+H_R]+\mathcal{L}_L[ Q]+\mathcal{L}_R[ Q],\label{Q}
\eea
where
\bea
H_L\equiv&&-\frac{i\hbar\gamma_L}{2}\sum_{\mu<\nu}\left(e^{ik|x_\mu-x_\nu|}\sigma_\mu^\dag\sigma_\nu-\textrm{H.c.}\right),\\
H_R\equiv&&-\frac{i\hbar\gamma_R}{2}\sum_{\mu>\nu}\left(e^{ik|x_\mu-x_\nu|}\sigma_\mu^\dag\sigma_\nu-\textrm{H.c.}\right),
\eea
denote the RDDI energy shifts, and the Lindblad forms,
\bea
\mathcal{L}_L[\hat Q]\equiv&&-\frac{\gamma_L}{2}\sum_{\mu,\nu}\Big\{e^{-ik(x_\mu-x_\nu)}\left(\sigma_\mu^\dag \sigma_\nu Q +Q \sigma_\mu^\dag\sigma_\nu \right.\nonumber\\
&&\left.-2\sigma_\mu^\dag Q\sigma_\nu\right)\Big\},\\
\mathcal{L}_R[\hat Q]\equiv&&-\frac{\gamma_R}{2}\sum_{\mu,\nu}\Big\{e^{ik(x_\mu-x_\nu)}\left(\sigma_\mu^\dag \sigma_\nu Q +Q \sigma_\mu^\dag\sigma_\nu \right.\nonumber\\
&&\left.-2\sigma_\mu^\dag Q\sigma_\nu\right)\Big\},
\eea
characterize the cooperative spontaneous decay under RDDI. The subscripts $L$ and $R$ respectively indicate the left- and right-propagating components of the 1D RDDI. We first note that the Lindblad forms here do not include non-guided couplings or other non-radiative losses, which could present in fibers or waveguides and reduce the overall efficiency of light collection. Furthermore, for atoms confined in 1D, we have ordering on the atomic positions, $x_1<x_2<...<x_{N-1}<x_N$, which otherwise does not present in the 2D and 3D cases, and we show in the below that this atomic ordering plays a role on cooperative spontaneous emissions.

When $\gamma_L=\gamma_R=\gamma$, we retrieve the usual reciprocal and infinite-range couplings of Eq. (\ref{chiral1D}),
\bea
J_{\mu,\nu}=\frac{\Gamma_{1D}}{2}\left[\cos(k_L x_{\mu,\nu})+i\sin(k_L |x_{\mu,\nu}|)\right],
\eea 
where $\Gamma_{1D}=2\gamma$ and $J_{\mu,\nu}=J_{\nu,\mu}$. Re[$J_{\mu,\nu}$] and Im[$J_{\mu,\nu}$] denote the incoherent and coherent parts of the couplings respectively. This infinite-range and cooperative dipole-dipole interaction in the 1D atom-fiber coupled system has been investigated theoretically \cite{Kien2005, Kien2008, Kien2017} and recently observed in macroscopically separated cold atoms \cite{Solano2017}. 

When single photon interacts with the atomic chain, only one of the atoms is excited. By defining 
\bea
F_{\mu\nu}\equiv&&\frac{\gamma_Re^{ik|x_{\mu,\nu}|}+\gamma_Le^{-ik|x_{\mu,\nu}|}}{2},\\
G_{\mu\nu}\equiv&&-i\frac{\gamma_Re^{ik|x_{\mu,\nu}|}-\gamma_Le^{-ik|x_{\mu,\nu}|}}{2},
\eea
and in terms of single excitation space $|\psi_\mu\rangle=|e\rangle_\mu|g\rangle^{\otimes(N-1)}$, we obtain the interaction Hamiltonian,
\begin{widetext}
\bea
V=&&\begin{bmatrix}
    -F_{11}  & -F_{12}+iG_{12} 		& -F_{13}+iG_{13} & \dots & -F_{1N}+iG_{1N} \\
    -F_{12}^*+iG_{12}^* & -F_{22} & -F_{23}+iG_{23}	& \dots & -F_{2N}+iG_{2N} \\
    -F_{13}^*+iG_{13}^* & -F_{23}^*+iG_{23}^* & -F_{33} & \dots & -F_{3N}+iG_{3N} \\
		\vdots 							& \vdots  & \vdots & \ddots & \vdots  \\
		-F_{1N}^*+iG_{1N}^* & -F_{2N}^*+iG_{2N}^* & -F_{3N}^*+iG_{3N}^* & \dots &-F_{NN}
\end{bmatrix}.
\eea
The atomic dynamics $|\Psi(t)\rangle=\sum_\mu c_\mu(t)|\psi_\mu\rangle$ is determined by the coupled equations: $\dot c_\mu=\sum_\nu V_{\mu,\nu}c_\nu$. In general, $F_{\mu\nu}$ and $G_{\mu\nu}$ are complex numbers, and $V$ is reciprocal, viz $V_{\mu,\nu}=V_{\nu,\mu}$, under $\gamma_L=\gamma_R=\gamma$.

By expressing $V$ as
\bea
V=&&\begin{bmatrix}
    -\frac{\gamma_L+\gamma_R}{2}  & -\gamma_Le^{-ik|x_{1,2}|} & -\gamma_Le^{-ik|x_{1,3}|} & \dots & -\gamma_Le^{-ik|x_{1,N}|}\\
    -\gamma_Re^{-ik|x_{1,2}|} & -\frac{\gamma_L+\gamma_R}{2} & -\gamma_Le^{-ik|x_{2,3}|} & \dots & -\gamma_Le^{-ik|x_{2,N}|}\\
    -\gamma_Re^{-ik|x_{1,3}|} & -\gamma_Re^{-ik|x_{2,3}|} & -\frac{\gamma_L+\gamma_R}{2} & \dots & -\gamma_Le^{-ik|x_{3,N}|}\\
		\vdots 							& \vdots  & \vdots & \ddots & \vdots  \\
		-\gamma_Re^{-ik|x_{1,N}|} & -\gamma_Re^{-ik|x_{2,N}|} & -\gamma_Re^{-ik|x_{3,N}|} & \dots & -\frac{\gamma_L+\gamma_R}{2}
\end{bmatrix},
\eea
\end{widetext}
the nonsymmetric feature of chiral coupling matrix emerges when $\gamma_R \neq \gamma_L$, and $V$ becomes nonreciprocal as $VV^\dag\neq V^\dag V$. This suggests that $V$ is not a normal matrix and cannot be unitarily diagonalized. Furthermore, it is a defective matrix which cannot be decomposed in terms of linearly independent eigenvectors. With this incomplete basis of eigenvectors for $V$ with nonreciprocal decay rates in general, eigen-decompositions do not apply in chiral-coupled systems. Therefore, we directly solve the system dynamics from 
\bea
\frac{d}{dt}\vec c=V\vec c,
\eea
where $\vec c\equiv [c_1(t),c_2(t),...,c_N(t)]$ with given initial conditions of $\vec c(t=0)$. Below we show analytical results for few-atom systems, and throughout the paper we consider uniform excitations of the atomic chain.

\subsection{Cascaded scheme}

First we investigate two and three atoms in the cascaded scheme \cite{Stannigel2012, Gardiner1993, Carmichael1993} where one of the nonreciprocal couplings $\gamma_{L,R}$ is zero, so that only unidirectional coupling is permitted. Starting with two atoms, we have the coupled equations,
\bea
\dot c_1(t)=&&-\frac{\gamma_L+\gamma_R}{2} c_1(t)-\gamma_L e^{-i\xi}c_2(t),\\
\dot c_2(t)=&&-\gamma_R e^{-i\xi}c_1(t)-\frac{\gamma_L+\gamma_R}{2} c_2(t),
\eea
where $\xi\equiv k|x_{1,2}|$, and correspondingly $\xi\lambda/(2\pi)$ represents the atomic separation $|x_{1,2}|$, given the transition wavelength $\lambda$. For reciprocal couplings where $\gamma_L=\gamma_R$ and assume $\xi=0$ or $2\pi$, we retrieve the conventional results of Dicke's super- and subradiance when $c_1(0)=1/\sqrt{2}$ and $c_2(0)=\pm 1/\sqrt{2}$, respectively. They give the symmetric and anti-symmetric states of single excitation spaces, $(|ge\rangle\pm|eg\rangle)/\sqrt{2}$. For an arbitrary $\xi$, with uniform excitations $c_1(0)=1/\sqrt{2}$, $c_2(0)=1/\sqrt{2}$, and considering the extreme case of nonreciprocal couplings in the cascaded scheme \cite{Stannigel2012} where $\gamma_L=0$ and $\gamma_R=\gamma$, we can solve for the above coupled equations,
\bea
c_1(t)=&&\frac{1}{\sqrt{2}}e^{-\gamma t/2},\label{N2}\\
c_2(t)=&&\frac{1}{\sqrt{2}}e^{-\gamma t/2} (1-\gamma t e^{-i\xi}).
\eea

For three atoms, we have the coupled equations, 
\bea
\dot c_1(t)=&&-\frac{\gamma_L+\gamma_R}{2} c_1(t)-\gamma_L e^{-i\xi}c_2(t)-\gamma_L e^{-i2\xi}c_3(t),\nonumber\\\label{N31}\\
\dot c_2(t)=&&-\gamma_R e^{-i\xi}c_1(t)-\frac{\gamma_L+\gamma_R}{2} c_2(t)-\gamma_L e^{-i\xi}c_3(t),\nonumber\\\label{N32}\\
\dot c_3(t)=&&-\gamma_R e^{-i2\xi}c_1(t)-\gamma_R e^{-i\xi}c_2(t)-\frac{\gamma_L+\gamma_R}{2} c_3(t),\nonumber\\\label{N33}
\eea
where uniform distributions of an atomic array is reflected on the phases $e^{-im\xi}$ with integers $m$. For the extreme case of $\gamma_L=0$, $\gamma_R=\gamma$, with the initial condition of uniform excitations $c_{1,2,3}(0)=1/\sqrt{3}$, we obtain
\bea
c_1(t)=&&\frac{e^{-\gamma t/2}}{\sqrt{3}},\label{N3}\\
c_2(t)=&&\frac{e^{-\gamma t/2} (1-\gamma t e^{-i\xi})}{\sqrt{3}},\\
c_3(t)=&&\frac{e^{-\gamma t/2} [\gamma^2 t^2 e^{-i2\xi}-2\gamma t(e^{-i\xi}+e^{-i2\xi})+2]}{2\sqrt{3}}.
\eea

\begin{figure}[t]
\centering
\includegraphics[width=8.5cm,height=4.5cm]{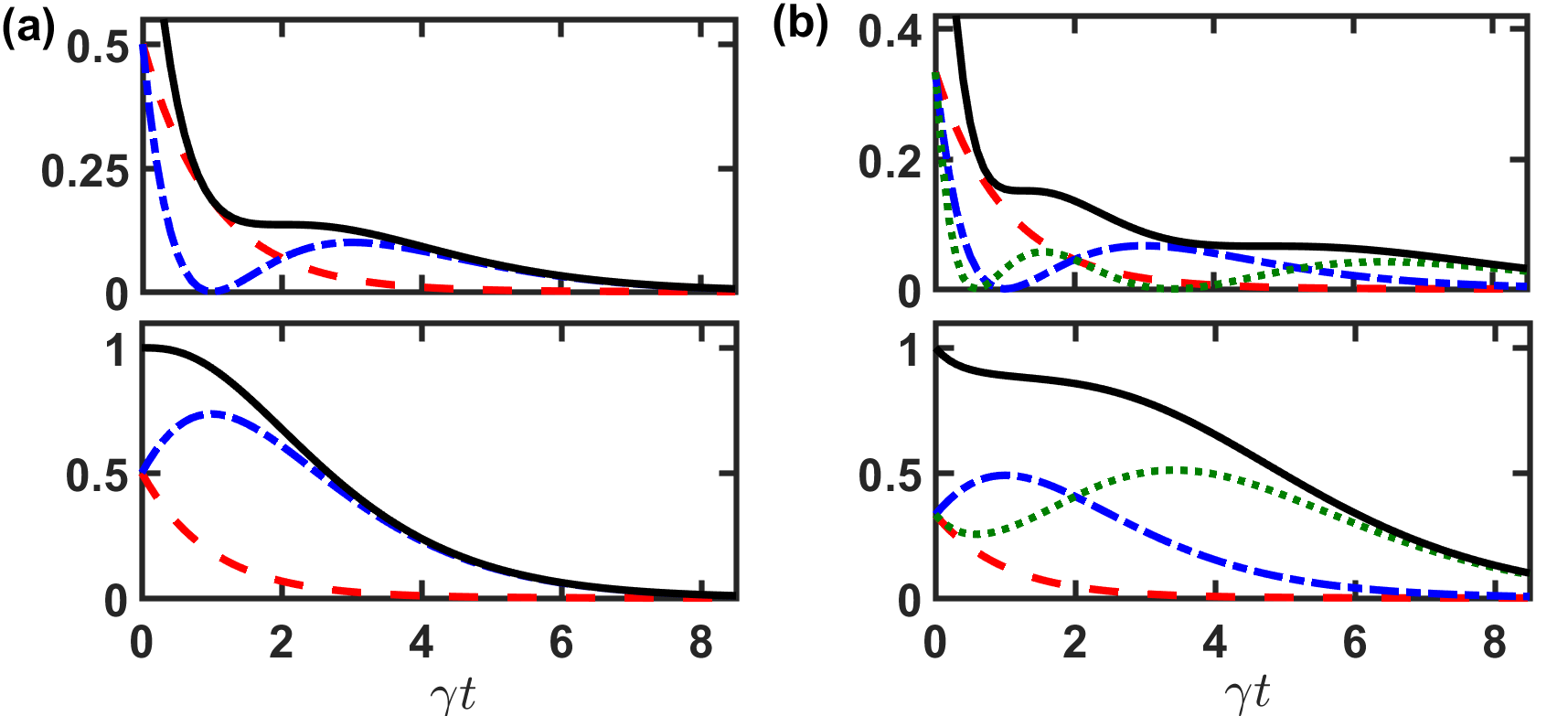}
\caption{Excited state populations of few atom systems for $\gamma_R=\gamma$ and $\gamma_L=0$. (a) For $N=2$, the upper and lower plots show the cases of $\xi$ $=$ $0$ and $\pi$ respectively. The excited state population of the first atom $P_1(t)$ (dashes in red) always decays as single atom $\propto e^{-\gamma t}$, while the second one $P_2(t)$ (dash-dots in blue) decays with repopulation and oscillation in either super- (upper plot) or subradiant rates (lower plot) in a short term. The total excited state population (solid line in black) shows nontrivial decay behaviors. (b) Excited state populations for $N=3$ with corresponding $\xi$'s in (a). In long time limit, the third atom (dots in green) decays slower than the second one (dash-dots in blue), while the second atom decays slower than the first one (dashes in red).}\label{fig2}
\end{figure}

In Fig. \ref{fig2}, we plot the excited state populations at specific $\xi$ in the cascaded scheme when $\gamma_R=0$, for two- and three-atom cases. As shown in Eqs. (\ref{N2}) and (\ref{N3}), the excited state population of the leftmost atom always decays as in the single atom (noninteracting) regime, which has a time dependence of $e^{-\gamma t}$, since there is no coupling to this atom from the atoms on the right. We find that at $\xi \lesssim \pi/4$, the total population $P_{tot}(t)=\sum_m P_m(t)\equiv |c_{m}(t)|^2$ shows an early superradiant decay followed by a subradiance, which can be seen in the upper panels of Fig. \ref{fig2}(a) and \ref{fig2}(b). The subradiant decay originates from the re-excitation of the atoms on the right by the (virtual) photon coming from the left, and this is deterministically achieved due to the light-induced correlations between the atoms. For even larger $\xi$ up to $\pi$, $P_{tot}(t)$ always decays subradiantly, and the repopulation emerges at different times depending on $\xi$. Specifically in the lowest plot of Fig. \ref{fig2}(b) for $\xi=\pi$, the atoms orderly repopulate the excited state and decay, i.e. $P_3(t)$ decays after $P_2(t)$. This indicates that the light excitation from the left can only transfer to the atoms on the right, blockading re-excitation of atoms on the left. This is also distinct for a 1D reservoir where light scattering and excitation exchange are allowed in one dimension only. When $\xi\approx\pi$, $P_{tot}(t)$ presents the most subradiant emission, which is reminiscent of the decoherence-free state in the setting with reciprocal couplings.

In such cascaded scheme ($\gamma_L=0$), only unidirectional coupling is allowed, and the re-excitation of the atoms on the right can be seen as the atoms in the setting of 1D reciprocal couplings with a perfect mirror on the left, which reflects the light leaving to the left back to the atoms. This is not possible for a conventional atomic chain without a waveguide, which scatters light in 3D free space. In the next subsection, we further investigate the non-cascaded scheme when $\gamma_L$ is finite.   

\subsection{Non-cascaded scheme}

\begin{figure}[t]
\centering
\includegraphics[width=8.5cm,height=4.5cm]{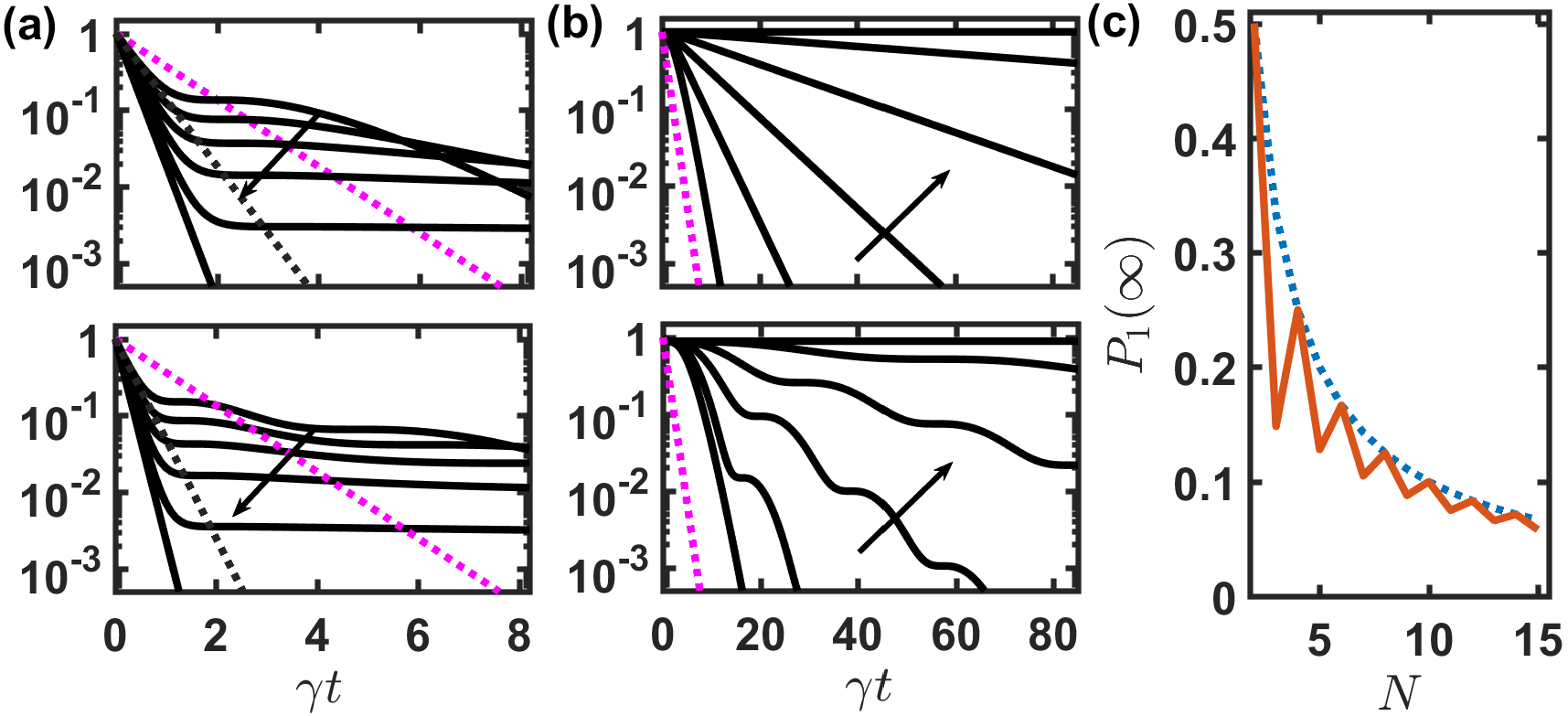}
\caption{Excited state populations in the cascaded and non-cascaded scheme. In (a) $\xi=0$ and (b) $\xi=\pi$, we plot $P_{tot}(t)$ for $N=2$ and $3$ in the upper and lower plots respectively. With a constant $\gamma_R=\gamma$, the arrows indicate the increasing $\gamma_L/\gamma=0$, $0.2$, $0.4$, $0.6$, $0.8$ to $1.0$. The dots in magenta give $e^{-\gamma t}$ and is plotted in (a) and (b) as references. The dots in black give $e^{-2\gamma t}$ and $e^{-3\gamma t}$ respectively in the upper and lower plots of (a) as comparisons with noninteracting regime. In (c) when $\gamma_L=\gamma_R$ at $\xi=\pi$, we show $P_1(t\rightarrow\infty)$ (solid), which approaches $N^{-1}$ (dotted) for odd $N$ asymptotically as $N$ increases.}\label{fig3}
\end{figure}

The non-cascaded scheme contrasts with the cascaded one when both left/right couplings are finite. For $N=2$ and $3$, we show $P_{tot}(t)$ in Fig. \ref{fig3} at two specific $\xi=0$ and $\pi$, which respectively show early superradiant and subradiant behaviors as in the cascaded scheme ($\gamma_L=0$). At $\xi=0$ in Fig. \ref{fig3}(a), both $P_{tot}(t)$ decay more superradiantly initially as $\gamma_L$ increases, which respectively approach $e^{-4\gamma t}$ and $e^{-6\gamma t}$ when $\gamma_L\rightarrow\gamma_R$, significantly faster than $e^{-2\gamma t}$ and $e^{-3\gamma t}$ in the noninteracting regime. This enhancement of superradiance is expected for uniform excitations and finite $\gamma_L$ at small $\xi$ or $\xi\sim 2\pi$, where all the atoms are in phase as in Dicke's superradiant regime. At much later time, these subradiant tails dissipate faster for smaller $\gamma_L$, which indicates of occupations, though fairly small, with longer decay time scales for $\gamma_L$ close to $\gamma_R$.  

In Fig. \ref{fig3}(b), we further show the subradiance dynamics at $\xi=\pi$ from cascade, to non-cascade, then to symmetric couplings. For both cases of $N=2$ and $3$, all atomic excitations prolong in time as $\gamma_L$ increases. This drives the system more subradiant toward decoherence-free states for even $N$, where $\gamma_L = \gamma_R$ and $P_m(\infty)=N^{-1}$. While for odd $N$ when $\gamma_L = \gamma_R$, the system becomes decoherence-free only when $N\rightarrow\infty$ with $P_m(\infty)\rightarrow N^{-1}$. This can be seen in Fig. \ref{fig3}(c) where we show $P_1(\infty)$ as an example. The total excited populations for odd $N$ never reaches one for finite $N$, and the remaining population should be in the ground state. This reflects that initially uniform excitation can never be decomposed in terms of decoherence-free eigenstates unless $N\rightarrow\infty$. As an example of $N=3$, the decoherence-free eigenstates are $\vec c=[-1,0,1]$ and $[1,1,0]$, whereas the third eigenstate is $[1,-1,1]$ with an eigenvalue of $-3\gamma$.   

In general for odd $N$ at $\xi=\pi$, the atoms on the edges decay faster, which can be seen in Eqs. (\ref{N31}) and (\ref{N33}) in the example of $N=3$, while they are repumped by the radiation of the central one. As time evolves, the central atom are excited and then decays, transferring the excitation more to the right than to the left given $\gamma_R>\gamma_L$. This leads to atomic population oscillations due to interferences of light transmissions and reflections. These exchanges of excitations play important roles in determining the radiation evolutions in the 1D chiral-coupled system, and manifest different decay behaviors for even and odd $N$, which we will discuss in the next section. Below we study longer atomic chain with nonreciprocal couplings and investigate the system especially for subradiant dynamics in longer time scales. As a final remark in this section, we note that Dicke's super- and subradiance under reciprocal couplings actually set the maximal and minimal bound of decay constants respectively. Therefore, the nonreciprocal couplings basically destroy (or partially destroy) the coherences required for Dicke's super- and sub-radiance. In other words, the super- and sub-radiant decay behaviors become less significant when the time-reversal symmetry in the couplings is broken. 

\section{Subradiance from a chiral-coupled atomic chain}

\begin{figure}[t]
\centering
\includegraphics[width=8.5cm,height=4.5cm]{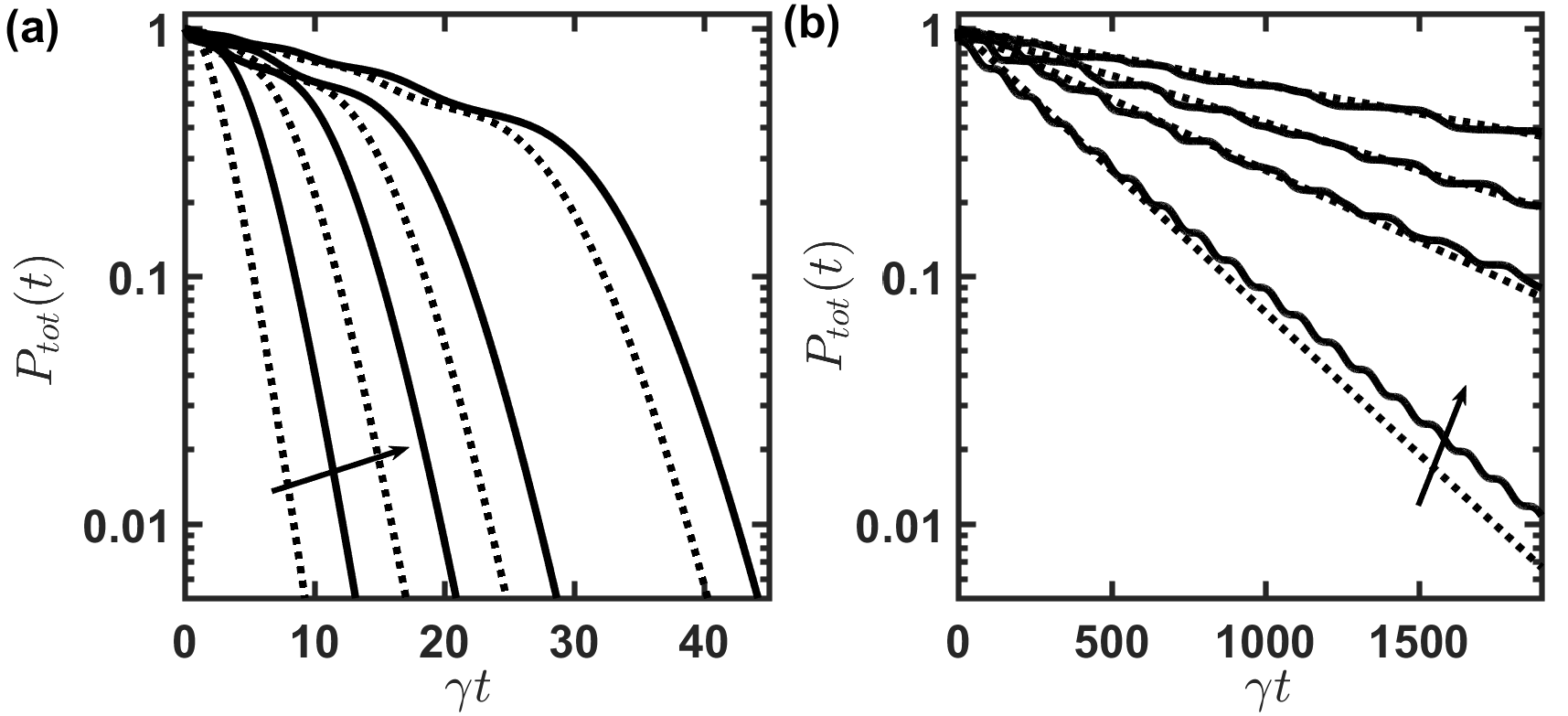}
\caption{Subradiance dynamics at $\xi=\pi$. The total atomic populations for (a) the cascaded case of $\gamma_L=0$ and (b) the non-cascaded case of $\gamma_L=0.9\gamma_R$ in logarithmic scales. The arrows indicate the increasing $N$ from $2-7$ to $10,11$ where we denote the even(odd) $N$ as dashed(solid) lines. The $P_{tot}$ shows clear plateaued regions of excitations for odd $N$ in (b).}\label{fig4}
\end{figure}

For a longer chiral-coupled atomic chain, we expect many-atom collective behaviors in the decay dynamics. In Fig. \ref{fig4}, we demonstrate the subradiance dynamics at $\xi=\pi$ as we increase $N$. In the cascaded scheme, the subradiant decay becomes more subradiant as $N$ increases. At least two decay time scales can be seen for the early and later stages of the exponential decay, which is evident in logarithmic plots, in contrast to the conventional exponential decay of noninteracting atoms. As $\gamma_L$ increases, we expect subradiance at a longer time as indicated in Fig. \ref{fig3}. As an example, we choose $\gamma_L$ close to $\gamma_R$ in Fig. \ref{fig4}(b), which presents a clear difference of subradiance between even and odd $N$. For odd $N$ in general, $P_{tot}(t)$ possesses excitation plateaus. This is due to temporally ordered atomic excitations as the system dissipates. The effect of smaller $\gamma_L$ modifies the overall decay, which has a shortened lifetime but still maintains the plateaued regions. In contrast to the subradiance of odd $N$, $P_{tot}(t)$ of even $N$ decays exponentially. This is due to the balanced excitation transfer between the atoms on even and odd sites. The ordered individual excitations from an odd $N$ atomic chain originates from the imbalanced population transfer, which can be seen in the early stage of the cases of $\xi=\pi$ in Figs. \ref{fig2}. The light-induced correlations between any two atoms in an odd $N$ atomic chain are further modified by the unpaired particle, which results in a $\pi$ phase change of the coherences $C_{nn'}(t)=c_n(t)c_{n'}^*(t)$, in contrast to the case of even $N$. Therefore, the plateaued excitations reflect this distinctive correlation in odd number of particles in the chiral-coupled chain, and can maintain in the subradiance dynamics. 

\begin{figure}[t]
\centering
\includegraphics[width=8.5cm,height=4.5cm]{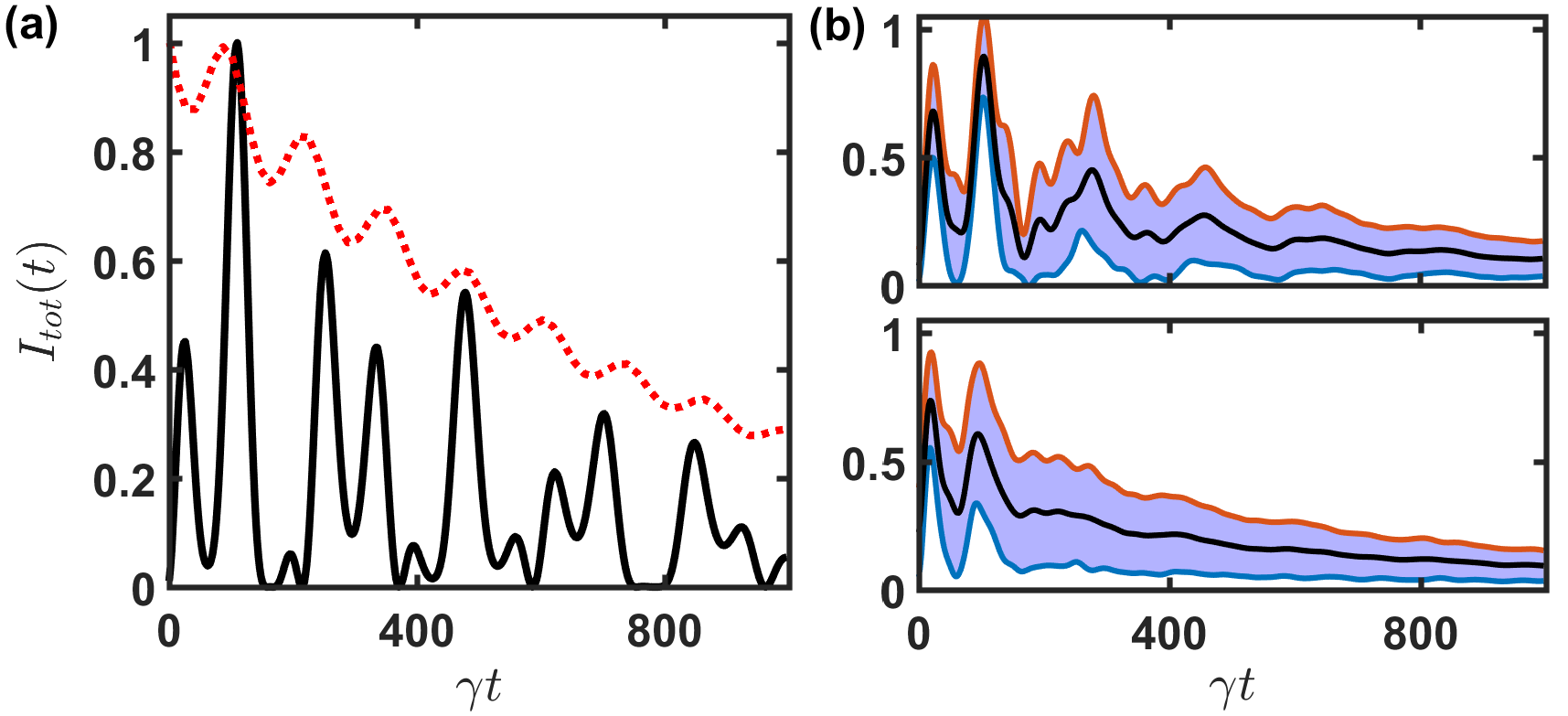}
\caption{Radiation dynamics at $\xi=\pi$ in the non-cascaded scheme of $\gamma_L=0.9\gamma_R$. (a) We show the time evolutions of emissions for the cases of $N=4$ (dashes in red) and $5$ (solid line in black) from the results of Fig. \ref{fig4}. (b) For $N=5$, we introduce position fluctuations in the upper and lower plots respectively with $0.5\%$ and $1\%$ randomly distributed deviations around the fixed positions. The shaded areas represent $1\sigma$ standard deviation of the mean curve (solid line in black) over ensemble averages.}\label{fig5}
\end{figure}

To compare with experimental observations, in Fig. \ref{fig5} we numerically calculate the radiation intensity, 
\bea
I_{tot}(t)=-\frac{dP_{tot}(t)}{dt}.\nonumber 
\eea
The burst emissions can be seen in Fig. \ref{fig5}, which reflects the clear plateaued regions of excitations in Fig. \ref{fig4}. For even $N$, the radiation evolution simply follows the exponential curve with small oscillations, in contrast to the bursts of radiation for odd $N$. The occurrences of the burst emission can, however, be reduced by position fluctuations. In Fig. \ref{fig5}(b), as the degree of fluctuation increases, the feature of burst radiation disappears. Since the chiral-coupled interactions are sensitively affected by the fluctuations of atomic positions, we expect that our predictions can be observable when the system experiences $<0.5\%$ of position fluctuations.

\section{Effect of atom dislocation}

In the above, we discussed subradiance dynamics in ideal conditions. Here we shall discuss non-ideal cases where dislocation of constituent atoms is present. This is necessary, because a precise positioning of the atoms is not easily fulfilled experimentally on the one hand, and on the other hand, disorder is known to induce Anderson localization. It is necessarily of interest to investigate the interplay between cooperative radiation and dislocation. Here the spatial variations of the atoms should be normalized to the transition wavelength. As such the effects of dislocation of superconducting qubits are less significant due to the long transmission wavelength, but for atoms with optical transitions, the effects are much more prominent. 

\begin{figure}[t]
\centering
\includegraphics[width=8.5cm,height=4.5cm]{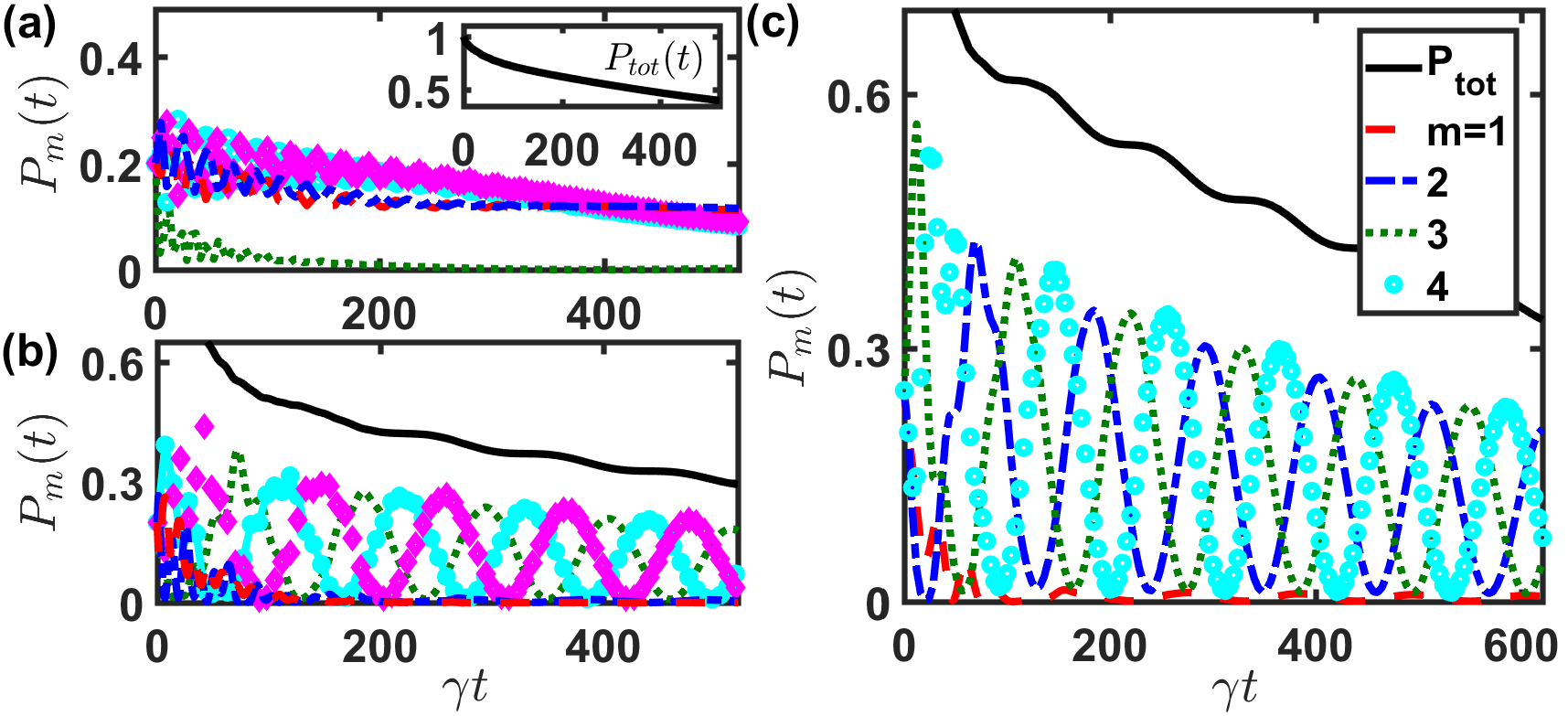}
\caption{Disorder-induced plateaued excitations. In the non-cascaded scheme as in Fig. \ref{fig4} with $N=5$, $\gamma_L=0.9\gamma_R$, and $\xi=\pi$, we place the spatial variation of $5\%$ to the right on (a) the central (the third) and (b) the second atoms. No plateaued excitation presents in (a), while plateaus reappear in (b). (c) Similarly for $N=4$, the leftmost atomic position is varied by $5\%$, and the plateaued excitation emerges, in contrast to Fig. \ref{fig4}. The respective line symbols for $P_m(t)$ are shown in the legend of (c), and $\diamond$ in magenta is for $m=5$.}\label{fig6}
\end{figure}

In Fig. \ref{fig6}, we show two examples of destruction and creation of plateaued excitations, where we add a spatial disorder at the level of a fraction of $\xi$. For odd number of atoms in Fig. \ref{fig6}(a), the central or edge disorder destroys the successive excitations that lead to the plateaued pattern when spatial variation $\gtrsim 3\%$. However, this can be restored when we place the disorder on the even sites for arbitrary spatial variations. This disorder can lead to initial fast decay either itself in Fig. \ref{fig6}(a) or with its neighboring atom in Fig. \ref{fig6}(b), which are negligibly small ($P_m(t)\lesssim 0.01$). For the spatial variation $\lesssim 2\%$, $P_{tot}(t)$ behaves as if no presence of the spatial deviation, and beyond which the disorder starts to make an effect on the radiation dynamics. On the other hand for even $N$ in Fig. \ref{fig6}(c), the plateaued excitation emerges due to the edge disorder $\gtrsim 2\%$, in contrast to Fig. \ref{fig4} where an atomic chain of even number decays exponentially without flattened regions. Clear plateaued regions can be seen and this indicates of spatially-dependent disorder-induced excitation plateaus in a chiral-coupled atomic chain, which allows a controllable way to manipulate the subradiant emission dynamics.

\begin{figure}[t]
\centering
\includegraphics[width=8.5cm,height=4.5cm]{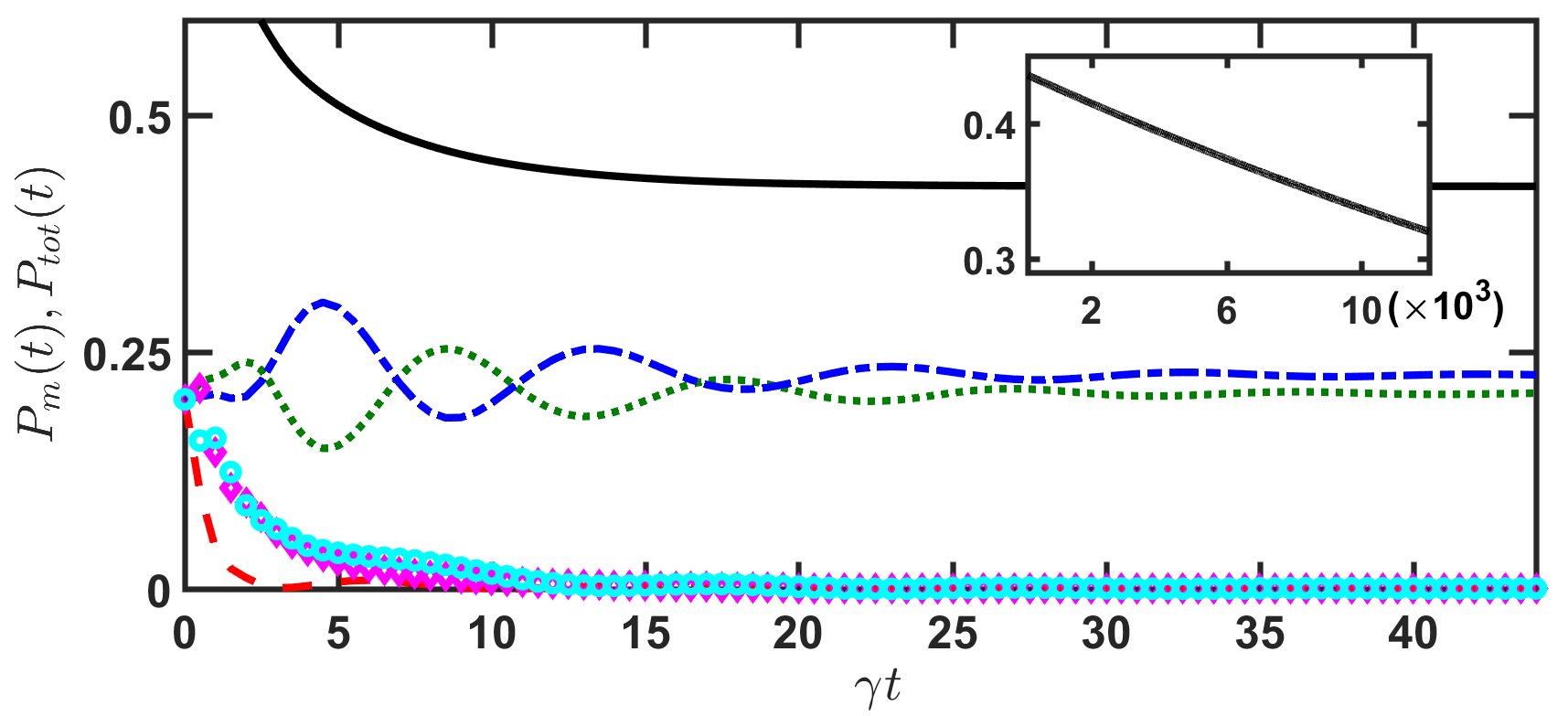}
\caption{Disorder-induced persistent localized excitations. In the non-cascaded scheme as in Fig. \ref{fig4} with $N=5$, $\gamma_L=0.9\gamma_R$, but here with $\xi=3\pi/4$ and $30\%$ spatial variation on the central atom, $P_{2,3}(t)$ sustains at finite populations for a very long time as shown in the inset. The line symbols are the same as in Fig. \ref{fig6}.}\label{fig7}
\end{figure}

Interestingly, in Fig. \ref{fig7}, when we choose $\xi\sim 3\pi/4$ and introduce a $\sim 30\%$ of position disorder on the $n$th non-edge atom, $P_n(t)$ and $P_{n-1}(t)$ preserve for a much longer time. This shows disorder-induced localized excitations, which can maintain up to $\gamma t\sim 10^4$ with only around $20\%$ reduction of the $P_{tot}(\gamma t\sim 100)$. Other parameter regimes, for example of $\xi=2.5\pi/4$ and $\sim 60\%$ disorder, can also support this localized excitation. This demonstrates a dimer-like excitation, which effectively forms a many-body state,
\bea
\left(\sqrt{1-P_{n-1}-P_n}+\sqrt{P_{n-1}}\sigma^\dag_{n-1} +\sqrt{P_{n}}\sigma^\dag_{n}\right)|g\rangle^{\otimes N},\nonumber
\eea
which can be prepared for very long time and controlled by local disorders. Moreover, for $n=2$ or $N$, $P_{n(n-1)}(t)$ decays much slower than the other $P_{m\neq n}(t)$, but not as the cases of $3\leq n\leq N-1$ which extend to a long time. This indicates the edge effect which involves the atoms at the boundary, where they decay to the left or right without back radiations. Nonetheless, the observation of these long-term behaviors can eventually be limited by the losses from non-guided modes.  

\section{Effect of imperfections of systems}

Here we discuss the effect of imperfections of chiral-coupled systems by including free space decays for each atoms.  This non-guided radiation loss can be present due to surface imperfections, scattering, and absorption of the waveguide or nanofiber, which leads to a 3D, non-chiral atom-photon coupling and poses a limitation on the results we have obtained. The inefficiency of guided mode couplings can be overcome, for example in an atom-fiber system, by aligning the atoms close to the nanofiber with an optimal fiber radius \cite{Kien2005_2, Dzsotjan2010, Chang2018} for a higher interaction strength. We treat the free space decay by including an extra one-body Lindblad term in Eq. (\ref{Q}),
\bea
\mathcal{L}_f[\hat Q]=-\frac{\gamma_f}{2}\sum_{\mu}(\sigma_\mu^\dag \sigma_\mu Q +Q \sigma_\mu^\dag\sigma_\mu -2\sigma_\mu^\dag Q\sigma_\mu),
\eea 
where $\gamma_f$ quantifies the effect of imperfections of chiral-coupled systems. This quantity has been measured in systems of an atom-waveguide \cite{Hung2013} and quantum dot in photonic-crystal waveguide \cite{Sollner2015, Arcari2014}, which is below one tenth of guided mode coupling and can be as low as $2\%$ of the directional coupling, respectively. While to fabricate homogeneous quantum dots are challenging, we note of the efforts to make scalable quantum dot arrays \cite{Zajac2016, Volk2019} by tuning quantum dots parameters \cite{Volk2019}, which can potentially simulate 1D topological phases \cite{Perez2019}. For atom-nanofiber setups, more than $90\%$ of optical power can be coupled to the desired directions \cite{Mitsch2014}, and in principle the non-guided radiation modes can be suppressed since it is proportional to $\sin(\xi)/\xi$ \cite{Solano2017}.

\begin{figure}[t]
\centering
\includegraphics[width=8.5cm,height=4.5cm]{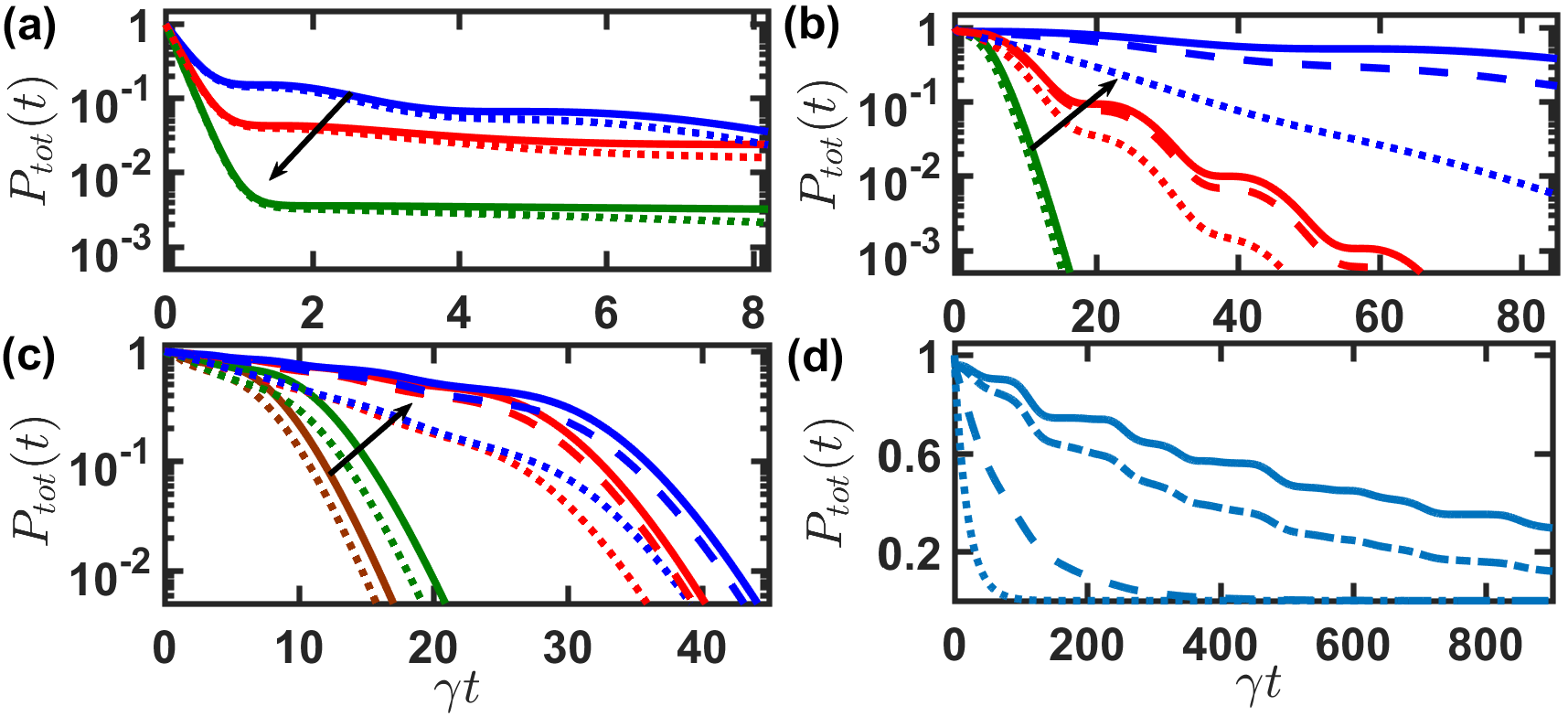}
\caption{Effect of imperfections of systems. With a constant $\gamma_R=\gamma$, we plot $P_{tot}(t)$ for $N=3$ at (a) $\xi=0$ and (b) $\xi=\pi$, where the arrows indicate the increasing $\gamma_L/\gamma=0$, $0.4$, and $0.8$ comparing to the case without $\gamma_f$ (solid) in Fig. \ref{fig3}. To compare Fig. \ref{fig4}, we consider $\xi=\pi$ for (c) $\gamma_L=0$ with the arrows indicating an increasing $N$ from $4,5,10$ to $11$, and for (d) $\gamma_L=0.9\gamma_R$ with $N=5$ as an example. We quantify the imperfection as $\gamma_f/\gamma=0.05$ (dots), $0.01$ (dashes), and $0.001$ (dash-dots). Various color lines in each subplots represent the cases under the same parameters with different $\gamma_f$. We have excluded some results of smaller $\gamma_f$ in (a), (b), and (c), since they are almost on top with the solid lines.}\label{fig8}
\end{figure}

In Fig. \ref{fig8}, we use a moderate $\gamma_f$ to demonstrate the effect of imperfections. For $\xi=0$ in Fig. \ref{fig8}(a), the non-guided radiation loss has no significant effect within a period of $1/\gamma_f$. This effect is augmented at a subradiant regime of $\xi=\pi$ in Fig. \ref{fig8}(b), where the feature of excitation plateaus becomes obscure when $\gamma_f$ increases. As for the multiatom effect, Fig. \ref{fig8}(c) shows a faster decay than the case without imperfections in Fig. \ref{fig4}. Nevertheless, we still can clearly find a multiple of time scales for the early and later stages of the decay. The effect of imperfection manifests the most near the decoherence-free condition when $\gamma_L$ is close to $\gamma_R$, which we show in Fig. \ref{fig8}(d) for $N=5$ as an example. The distinct excitation plateaus can barely be retrieved unless we make $\gamma_f$ small enough. To experimentally observe the subradiance dynamics in chiral-coupled systems, a timescale of $1/\gamma_f$ can be a good estimate of time window, within which the clear signature of it is allowed. Furthermore, $\gamma_f$ relates to $\beta=(\gamma_L+\gamma_R)/(\gamma_L+\gamma_R+\gamma_f)$ factor \cite{Arcari2014,Tiecke2014}, which is a ratio between the rate of spontaneous emissions into the guided modes and the total emission rate of all modes \cite{Lodahl2017}. 
This factor also corresponds to the single-emitter cooperativity $\eta=(\gamma_L+\gamma_R)/\gamma_f$, which is required to be $\eta\gg 1$ for a genuine quantum phase switch \cite{Tiecke2014}, for example. This condition is similar for the subradiance dynamics studied here to be clearly observed in realistic experiments, where strong atom-photon couplings are required. For the examples in Figs. \ref{fig8}(b) and \ref{fig8}(c), the cases of $\gamma_L/\gamma=0.4$ (red dots for $\gamma_f/\gamma=0.05$) and $0$ (red dashes for $\gamma_f/\gamma=0.01$) correspond to $\beta=96.6\%$ and $99\%$, respectively, which still show clear features of plateaued excitations and two-time decay curves under moderate radiation losses and are within reach of nowadays experiments of $\beta= 98\%$ \cite{Sollner2015, Arcari2014}. As an estimate for a time window of $\gamma t\sim 200=1/(\gamma_f/\gamma)$ for longer time dynamics of subradiance, $\beta\sim 99.5\%$ and $\eta\sim 200$ in the cascaded scheme where $\gamma_R=\gamma$ and $\gamma_L=0$. This demanding strong coupling regime however can be assisted and enhanced by an external cavity. 

An alternative approach to reach strong coupling regime is to place quantum dots on an optical nanofiber \cite{Yala2014}, which can enhance spontaneous emissions into the guided modes in this on-fiber light-matter interface \cite{Yala2014, Solano2017_2}. For the moment, an atom-nanofiber system is under fast development and investigated excessively. We note of a recent progress on single collective excitation in an atom-nanofiber system \cite{Corzo2019}, which shows a capability for generating non-classical quantum states and is essential for waveguide quantum electrodynamics. Therefore, we expect a larger $\beta$ factor promising in the near future.   

\section{Conclusion}

In conclusion, we have investigated the subradiant excitations from a chiral-coupled 1D atomic chain, which allows an infinite-range dipole-dipole interaction and nonreciprocal radiation coupling. The subradiance arisen from this non-conventional coupling shows rich dynamics. In this chiral-coupled 1D system where the right and left decay channels are finite but different, we can initiate superradiance or subradiance depending on the inter-atomic spacing in a uniformly distributed atomic chain. When the number of the atoms increases, the subradiant decay rate decreases, which indicates multi-atom enhancement. This non-cascaded scheme also allows sequential radiations from the ordered atoms, which form a series of excitation plateaus due to the violation of the time reversal symmetry, in contrast to the case of reciprocal couplings where $m$th and $(N-m+1)$th atoms should behave exactly the same in time. This feature of burst emissions following plateaued excitations is hindered by fluctuations of atomic positions, but can be sustained and observable as long as the fluctuations are kept small.   

Furthermore, by introducing local disorder in space and strength, the excitation plateaus can be tuned to appear or disappear. We also obtain the disorder-induced localized excitations which can maintain for a very long time. This suggests dynamical dimer-like state components spontaneously emerged from the system via dissipation. An even richer dynamics can be possible for multiple disorders or multi-photon excitations, such that potentially trimerized or complex correlations can emerge in the chiral-coupled system. For realistic considerations, we include a non-guided radiation loss, which puts a limit of time window on the results we obtain. This loss however can be kept small when strong atom-photon couplings are achieved. Finally, our investigations on subradiance is of interest to quantum storage of guided emissions \cite{Sayrin2015}. This can be done, for example in superconducting qubit systems, by preparing subradiant states from superradiant ones with controllable single-qubit phase gates \cite{Wang2020}. We also expect potential applications in many-body spin dynamics \cite{Hung2016} with chiral and infinite-range couplings, which can lead to interaction-driven phases of crystalline orders or bi-edge (hole) excitations in driven-dissipative chiral-coupled systems \cite{Jen2020_PRR}, or many-body localization dynamics in superconducting qubits \cite{Xu2018}.

\section*{Acknowledgements}

This work is supported by the Ministry of Science and Technology (MOST), Taiwan, under the Grant No. MOST-106-2112-M-001-005-MY3, 106-2811-M-001-130, and 107-2811-M-001-1524. GDL thanks the support from MOST of Taiwan under Grant No. 105-2112-M-002-015-MY3 and National Taiwan University under Grant No. NTU-CC-108L893206. We are also grateful for the support of NCTS ECP1 (Experimental Collaboration Program).

\appendix*
\section{Cooperative spontaneous emissions in one- and two-dimensional reservoirs}
\subsection{General formalism}
Cooperative spontaneous emissions in a three-dimensional (free space) reservoir has been investigated \cite{Lehmberg1970}, where resonant dipole-dipole interaction (RDDI) in the dissipation process emerges due to the common light fields mediating the whole atomic system. This pairwise dipole-dipole interaction underlies the super- and subradiance which hugely depend on the atomic spatial configurations. These contrasted fast and slow decay phenomena, especially from a dense medium, are originated from nonclassical many-body states \cite{Jen2016_SR} accessible in a light-matter interacting system.

Following the line of deriving RDDI in a three-dimensional (3D) reservoir \cite{Lehmberg1970}, here we review the cooperative spontaneous emissions in one- \cite{Tudela2013} (1D) and two-dimensional (2D) reservoirs. From a system of $N$ two-level atoms ($|g\rangle$ and $|e\rangle$ for the ground and excited states respectively) interacting with a reservoir composed of quantized bosonic fields, we have the Hamiltonian with a dipole approximation,
\bea
H =&& \sum_{\mu=1}^N \hbar\omega_e\hat\sigma_\mu^\dag\hat\sigma_\mu-\sum_{\mu=1}^N\sum_q g_q (e^{i\k_q\cdot\r_\mu-i\omega_q t}\hat a_q\nonumber\\
&&+e^{-i\k_q\cdot\r_\mu+i\omega_q t}\hat a_q^\dag) (\hat \sigma_\mu +\hat\sigma_\mu^\dag),
\eea
where $\hat\sigma_\mu\equiv|g\rangle_\mu\langle e|$, and bosonic fields $\hat a_q$ should satisfy commutation relations $[\hat a_q,\hat a_{q'}^\dag]=\delta_{q,q'}$. Note that the above light-matter interaction involves non-rotating wave terms. These often neglected terms (called rotating-wave approximation which is valid since twice the optical frequency of the interacting light field is averaged out) are crucial for a complete description of the frequency shift (dispersion) in the dissipation, which therefore should satisfy the Kramers-Kronig relation with the decay rate (absorption). The coupling constant is $g_q\equiv d/\hbar\sqrt{\hbar\omega_q/(2\epsilon_0V)}(\vec\epsilon_q\cdot\hat d)$ with a dipole moment $d$, unit direction of the dipole $\hat d$, field polarization $\vec\epsilon_q$, and quantization volume $V$.

Consider a Heisenberg equation for an atomic operator $\hat Q$, we have $d\hat Q/dt=i[H,\hat Q]$ (setting $\hbar=1$), which reads
\bea
\frac{d\hat Q}{dt}=&&i\omega_e\sum_\mu[\hat\sigma_\mu^\dag\hat\sigma_\mu,\hat Q]-i\sum_\mu\sum_q g_q\{e^{i\k_q\cdot\r_\mu}[\hat\sigma_\mu+\hat\sigma_\mu^\dag,\hat Q]\nonumber\\
&&\times\hat a_q(t)-e^{-i\k_q\cdot\r_\mu}\hat a_q^\dag(t)[\hat Q,\hat\sigma_\mu+\hat\sigma_\mu^\dag]\}.
\eea
The above involves $\hat a_q$ which can be further solved from $d\hat a_q/dt=i[H,\hat a_q]$. The solution of $\hat a_q$ reads
\bea
\hat a_q(t)=&&\hat a_q(0)e^{-i\omega_q t}+i\sum_\mu g_qe^{-i\k_q\cdot\r_\mu}\int_0^t dt' [\hat\sigma_\mu(t')\nonumber\\&&+\hat\sigma_\mu^\dag(t')]e^{-i\omega_q(t-t')}.
\eea
With Born-Markov approximation, equivalently considering a relevant dynamical timescale of $\omega_e t\gg 1$ and $t\gg (r_{\mu\nu})_{max}/c$ ($r_{\mu\nu}\equiv |\r_\mu-\r_\nu|$) \cite{Lehmberg1970}, we derive the dynamical Heisenberg equations of $Q\equiv\langle\hat Q\rangle_0$ in Lindblad forms by assuming the vacuum initial bosonic fields $\langle\rangle_0$ (equivalent to the trace of the bosonic fields, leading to the reduced density matrix equations of the atoms),
\bea
\dot{Q}(t)&&=\sum_{\mu\neq\nu}i\Omega_{\mu,\nu}[\sigma_\mu^\dag\sigma_\nu,Q]+\mathcal{L}(Q),\\
\mathcal{L}(Q)&&=\sum_{\mu,\nu}\gamma_{\mu,\nu}\left[\sigma_\mu^\dag Q\sigma_\nu-\frac{1}{2}(\sigma_\mu^\dag\sigma_\nu Q+Q\sigma_\mu^\dag\sigma_\nu)\right].
\eea 
The pairwise couplings $(\gamma_{\mu,\nu}+i2\Omega_{\mu,\nu})/2\equiv J_{\mu,\nu}$ which can be defined as
\begin{widetext}
\bea
J_{\mu,\nu}&&=\sum_q |g_q|^2\int_0^\infty dt' e^{i\k_q\cdot(\r_\mu-\r_\nu)}[e^{i(\omega_e-\omega_q)t'}+e^{-i(\omega_e+\omega_q)t'}], \nonumber\\
&&=\sum_q |g_q|^2\int_0^\infty dt' e^{i\k_q\cdot(\r_\mu-\r_\nu)}[\pi\delta(\omega_q-\omega_e)+\pi\delta(\omega_q+\omega_e)+i\mathcal{P}(\omega_e-\omega_q)^{-1}-i\mathcal{P}(\omega_q+\omega_e)^{-1}].\label{J}
\eea

For a 3D reservoir, we let $\sum_q\rightarrow\sum_{\vec\epsilon_q}\int_{-\infty}^\infty\frac{V}{(2\pi)^3}d^3q$ with two possible field polarizations $\vec\epsilon_q$. In spherical coordinates, we show the main results of $J_{\mu,\nu}$ in free space \cite{Lehmberg1970},
\bea
\gamma_{\mu,\nu}(\xi)&&\equiv\oint d\Omega_q[1-(\hat\q\cdot\hat\p)^2]\int_0^\infty dq q^2 \bar g_q^2\frac{V}{(2\pi)^3}[\pi\delta(\omega_q-\omega_e)+\pi\delta(\omega_q+\omega_e)],\nonumber\\
&&=\frac{3\Gamma}{2}\bigg\{\left[1-(\hat\p\cdot\hat{r}_{\mu\nu})^2\right]\frac{\sin\xi}{\xi}
+\left[1-3(\hat\p\cdot\hat{r}_{\mu\nu})^2\right]\left(\frac{\cos\xi}{\xi^2}-\frac{\sin\xi}{\xi^3}\right)\bigg\},\label{F}\\
\Omega_{\mu,\nu}(\xi)&&\equiv-\oint d\Omega_q[1-(\hat\q\cdot\hat\p)^2]\int_0^\infty dq q^2 \bar g_q^2\frac{V}{(2\pi)^3}[i\mathcal{P}(\omega_q-\omega_e)^{-1}+i\mathcal{P}(\omega_q+\omega_e)^{-1}],\nonumber\\
&&=\frac{3\Gamma}{4}\bigg\{-\Big[1-(\hat\p\cdot\hat{r}_{\mu\nu})^2\Big]\frac{\cos\xi}{\xi}
+\Big[1-3(\hat\p\cdot\hat{r}_{\mu\nu})^2\Big]
\left(\frac{\sin\xi}{\xi^2}+\frac{\cos\xi}{\xi^3}\right)\bigg\}\label{G}, 
\eea
\end{widetext}
where $d\Omega_q$ denotes an integration of a solid angle of $4\pi$, $\mathcal{P}$ is the principal value of the integral, $\bar g_q^2$ $\equiv$ $(d/\hbar)^2[\hbar\omega_q/(2\epsilon_0V)]$, $\hat\p$ aligns with the excitation field polarization, the intrinsic decay constant $\Gamma=d^2\omega_e^3/(3\pi\hbar\epsilon_0c^3)$, and $\xi\equiv k_L|\r_\mu-\r_\nu|$ with $k_L=\omega_e/c$. $\gamma_{\mu,\nu}$ and $\Omega_{\mu,\nu}$ are respectively collective decay rates and frequency shifts in general for any two atoms in the ensemble. As $\xi\rightarrow 0$, Dicke's regime is reached where $\gamma_{\mu,\nu}\rightarrow \Gamma$, while $\Omega_{\mu,\nu}$ diverges. In the below, we review the results of $J_{\mu,\nu}$ in one- and two-dimensional reservoirs. 

\subsection{One-dimensional reservoir}

From Eq. (\ref{J}), the 1D reservoir gives \cite{Tudela2013} (note that $V$ in $\bar g_q$ is changed to $L$ as a length scale of one-dimensional quantization volume)
\bea
J_{\mu,\nu}=&&\int_{-\infty}^\infty\frac{dq}{2\pi}\bar g_q^2 L e^{i\k_q\cdot(\r_\mu-\r_\nu)}[\pi\delta(\omega_q-\omega_e)\nonumber\\&&+\pi\delta(\omega_q+\omega_e)+i\mathcal{P}(\omega_e-\omega_q)^{-1}-i\mathcal{P}(\omega_q+\omega_e)^{-1}].\nonumber\\
\eea
We further obtain (let $x_{\mu,\nu}=x_\mu-x_\nu$ and dropping $q$ in $\omega_q$ for brevity)
\bea
J_{\mu,\nu}=&&\int_0^\infty\frac{d\omega}{\pi}|\partial_\omega q(\omega)|\bar g_q^2 L\cos(\k_qx_{\mu,\nu})[\pi\delta(\omega-\omega_e)\nonumber\\&&+\pi\delta(\omega+\omega_e)+i\mathcal{P}(\omega_e-\omega)^{-1}-i\mathcal{P}(\omega+\omega_e)^{-1}].\nonumber\\
\eea
Let $\Gamma_{1D}\equiv 2|\partial_\omega q(\omega)|_{\omega=\omega_e}\bar g_{k_L}^2L$, where we keep the dispersion relation of the 1D coupling constant, and we obtain
\bea
J_{\mu,\nu}=&&\frac{\Gamma_{1D}}{2}\cos(k_L x_{\mu,\nu})-\frac{i\mathcal{P}}{\pi}\int_{-\infty}^\infty d\omega\nonumber\\&&\times
\frac{{\rm Re}[|\partial_\omega q(\omega)|\bar g_q^2Le^{i\k_q(x_\mu-x_\nu)}]}{\omega-\omega_e},
\eea
which shows the Kramers-Kronig relation between real and imaginary parts of $J_{\mu,\nu}$. Finally the cooperative spontaneous emissions in 1D reservoir becomes \cite{Tudela2013}
\bea
J_{\mu,\nu}=\frac{\Gamma_{1D}}{2}\left[\cos(k_L x_{\mu,\nu})+i\sin(k_L |x_{\mu,\nu}|)\right].\label{chiral1D}
\eea 
The above sinusoidal form shows the infinitely long-range dipole-dipole interaction between any atoms. This interaction, in contrast to 3D and later 2D results below, does not diverge in the part of collective frequency. Moreover, a true Dicke regime of $J_{\mu,\nu}=\Gamma_{1D}/2$ (no dipole-dipole interaction energy shift) when $\xi=0,~2\pi$ can be realized in 1D light-matter interacting system, making such system a potentially highly dynamical and strongly interacting platform. Meanwhile, when $\xi=\pi/2,~3\pi/2$, RDDI in 1D reservoir allows only intrinsic decay and coherent exchange between atoms without dissipation. 

\subsection{Two-dimensional reservoir}

Similarly from Eq. (\ref{J}), the 2D reservoir has a quantization volume $A$, and we obtain $J_{\mu,\nu}$ in polar coordinates,
\begin{widetext}
\bea
J_{\mu,\nu}=\int_0^\infty\frac{q\bar g_q^2 A}{(2\pi)^2}  dq\int_0^{2\pi} d\theta [1-(\hat \q\cdot\hat\p)^2]e^{i\k_q\cdot\r_{\mu,\nu}}[\pi\delta(\omega-\omega_e)+\pi\delta(\omega+\omega_e)+i\mathcal{P}(\omega_e-\omega)^{-1}-i\mathcal{P}(\omega+\omega_e)^{-1}].
\eea
Consider the real part of $J_{\mu,\nu}$ first and $\hat\q$ has polar angle $\theta$ to the $\hat z$. We assume that $\r_{\mu,\nu}$ is along $\hat z$ and $\hat\p_\parallel$ aligns with a polar angle $\theta'$ to $\hat z$ where $\hat\p_\parallel=\hat\p\sin\phi$ and $\hat\p_\perp=\hat\p\cos\phi$ with an angle $\phi$ to $\hat y$. We then obtain
\bea
{\rm Re}[J_{\mu,\nu}]=&&\int_0^\infty\frac{q\bar g_q^2 A}{(2\pi)^2}  dq\int_0^{2\pi} d\theta[1-(\cos\theta\cos\theta'+\sin\theta\sin\theta')^2\sin^2\phi]e^{i\xi\cos\theta}\pi\delta(\omega-\omega_e),\nonumber\\
=&&\frac{\Gamma_{2D}}{2}\frac{1}{\pi}\int_0^{2\pi} d\theta[1-(\cos\theta\cos\theta'+\sin\theta\sin\theta')^2\sin^2\phi]e^{i\xi\cos\theta}
\eea 
\end{widetext}
where $\Gamma_{2D}\equiv 2k_L|\partial_\omega q(\omega)|_{\omega=\omega_e}\bar g_{k_L}^2A\pi^2/(2\pi)^2$.

To further calculate ${\rm Re}[J_{\mu,\nu}]$, we need the following integrals,
\bea
&&\int_0^{2\pi}e^{ia\cos\theta}d\theta=2\pi J_0(|a|),\\
&&\int_0^{2\pi}\cos^2\theta e^{ia\cos\theta}d\theta=2\pi (\frac{J_1(a)}{a}-J_2(a)),\\
&&\int_0^{2\pi}\sin^2\theta e^{ia\cos\theta}d\theta=2\pi \frac{J_1(|a|)}{|a|},\\
&&\int_0^{2\pi}\sin\theta\cos\theta e^{ia\cos\theta}d\theta=0,
\eea
where $J_n(a)$ represents the Bessel functions of the first kind. Then we obtain
\bea
{\rm Re}[J_{\mu,\nu}]=&&\frac{\Gamma_{2D}}{2}2\bigg[J_0(\xi)-\frac{J_1(\xi)}{\xi}+(\hat\p\cdot\hat\r_{\mu,\nu})^2J_2(\xi)\bigg],\nonumber\\
\equiv&&\frac{\Gamma_{2D}}{2}f(\xi).
\eea
The ${\rm Re}[J_{\mu,\nu}]$ and ${\rm Im}[J_{\mu,\nu}]$ should satisfy the Kramers-Kronig relation, and we obtain
\bea
J_{\mu,\nu}=&&\frac{\Gamma_{2D}}{2}f(\xi)-\frac{i\mathcal{P}}{2\pi}\int_{0}^\infty d\omega\left(\frac{1}{\omega-\omega_e}+\frac{1}{\omega+\omega_e}\right)
\nonumber\\&&\times(k_L|\partial_\omega q(\omega)|\bar g_q^2A/2)f(\omega|\r_\mu-\r_\nu|/c),\nonumber\\
=&&\frac{\Gamma_{2D}}{2}[f(\xi)+ig(\xi)],
\eea
where 
\bea
f(\xi)\equiv&& 2\left[J_0(\xi)-\frac{J_1(\xi)}{\xi}+(\hat\p\cdot\hat\r_{\mu,\nu})^2J_2(\xi)\right],\\
g(\xi)\equiv&& 2Y_0(\xi)-2\frac{Y_1(\xi)}{\xi}+2(\hat\p\cdot\hat\r_{\mu,\nu})^2Y_2(\xi)\nonumber\\&&
-\frac{4}{\pi\xi^2}[1-2(\hat\p\cdot\hat\r_{\mu,\nu})^2],
\eea
and $Y_n(\xi)$ represents the Bessel functions of the second kind. The above $g(\xi)$ can be derived by using the following integrals,
\bea
&&\mathcal{P}\int_0^\infty da\frac{J_0(a)}{a\mp b}=-\frac{\pi}{2}[Y_0(b)\pm H_0(b)],\\
&&\mathcal{P}\int_0^\infty da\frac{J_1(a)}{a(a\mp b)}=-\frac{2+\pi b[Y_1(b)\pm H_1(b)]}{2b^2},\\
&&\mathcal{P}\int_0^\infty da\left(\frac{J_2(a)}{a- b}+\frac{J_2(a)}{a+ b}\right)=-\frac{4}{b^2}-\pi Y_2(b),
\eea
where $H_n(b)$ is the Struve function.

All the results for the RDDI in 1D, 2D, and 3D reservoirs are reciprocal. This means that $J_{\mu,\nu}=J_{\nu,\mu}$, preserving the time reversal symmetry for light scattering between the $\mu$th and $\nu$th atoms. Similar 2D RDDI has been studied in point scatterers using 2D coupled dipole equations \cite{Maximo2015}, and super- and subradiance properties are investigated in details with the above 2D RDDI \cite{Jen2019_2D}.

\end{document}